%% file: FB.tex
\documentclass{PoS}
\usepackage{amsmath,epsfig,nico}

\title{Heavy-light decay constant at the $1/m$ order of HQET}

\ShortTitle{Heavy-light decay constant at the $1/m$ order of HQET}

\author{\epsfxsize=2.5 true cm
        \epsfbox{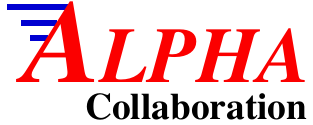} \vspace{-1.8cm} \hfill 
\onecol{4.5cm}{\vspace{-1cm}\it DESY 07-160 \\ SFB/CPP-07-56 \\  CERN-PH-TH/2007-172}
}

\author{Beno\^it Blossier \thanks{
This work has been supported by the Deutsche
Forschungsgemeinschaft (DFG) in SFB Transregio 9
``Computational Particle Physics''.}\\
  DESY,
  Platanenallee 6,
  15738 Zeuthen,
  Germany }

\author{Michele Della Morte \\
  CERN, Physics Department, TH Unit, 
  CH-1211 Geneva 23, 
  Switzerland }

\author{\speaker{Nicolas Garron}\\ 
  DESY,
  Platanenallee 6,
  15738 Zeuthen,
  Germany \\
  E-mail: \email{Nicolas.Garron@Desy.de}}

\author{Rainer Sommer \\
  DESY,
  Platanenallee 6,
  15738 Zeuthen,
  Germany }

\abstract{
Following the strategy developed by the ALPHA collaboration, we present a method
to compute non-perturbatively the decay constant of a heavy-light meson in
HQET including the 1/m corrections. We start by a matching between HQET and
QCD in a small volume to determine the parameters of the effective theory
non-perturbatively. Observables in the effective theory are then evolved to
larger volumes. In two steps a large enough volume is reached to determine the
physical decay constant. Some preliminary results in the quenched
approximation are shown.}

\FullConference{The XXV International Symposium on Lattice Field Theory\\
		 July 30-4 August 2007\\
		 Regensburg, Germany}

\begin{document}

\section{Introduction}
A few years ago, a non-perturbative formulation of Heavy Quark Effective
Theory (HQET) has been given
in~\cite{Heitger:2003nj} - see~\cite{DellaMorte:2007} 
for a review given at this conference. In particular the
problem of power divergences is solved through a finite volume matching.
Last year, this has been applied to the quenched computation of the b-quark
mass at the $\minv$ order~\cite{DellaMorte:2006cb}. In the same spirit, we present
here a strategy to compute a heavy-light decay constant.
We start by writing the Lagrangian at the leading order (i.e. in the static
approximation) and add a kinetic and a
magnetic piece (we follow the conventions of~\cite{DellaMorte:2006cb} 
and set the counterterm $\delta_m$ to zero)
\be
\lag{HQET} = \heavyb \,D_0\, \heavy 
- \omegakin  \heavyb{\bf D}^2\heavy \,
- \omegaspin \heavyb{\boldsymbol\sigma}\!\cdot\!{\bf B}\heavy \; .
\ee
A precise definition of the operators $D_0$, ${\bf D}^2$ 
and ${\boldsymbol\sigma}\!\cdot\!{\bf B}$ can be found 
in~\cite{DellaMorte:2006cb}. Here we just note that $\omegakin$ 
and $\omegaspin$ are some bare parameters of the effective theory.
\subsection{Schr\"odinger functional (SF) correlation functions}

In QCD, we consider the (renormalized and improved) current to boundary
correlators 
$\fa$ and $k_V$ 
defined - up to improvement factors such as $(1+b_{\rm A} a m_{\rm q,b})$ - in the SF by 
\bea
\fa(x_0) &=& -Z_A \, Z_\zeta^2 \, {a^6 \over 2}\, \sum_{{\bf y}, {\bf z}}
\la \left(A_{\rm I}\right)_0 (x) \, \bar \zeta_{\rm b}({\bf y}) \gamma_5 \zeta_{\rm l}({\bf z}) \ra, \\
\kv(x_0) &=& -Z_V \, Z_\zeta^2 \, {a^6 \over 6} \sum_{{\bf y}, {\bf z}, k} 
\la \left(V_{\rm I}\right)_k (x) \, \bar \zeta_{\rm b}({\bf y}) \gamma_k
\zeta_{\rm l}({\bf z})  \ra, 
\eea
where the improved currents  $A_{\rm I}(x)$ and  $V_{\rm I}(x)$ are defined as in \cite{Heitger:2003nj}. \\
We also consider the boundary to boundary correlators 
\bea
\fone &=& -Z_\zeta^4 \, {a^{12} \over 2L^6} \, \sum_{{\bf u}, {\bf v}, {\bf y}, {\bf z}}
\la \bar \zeta_{\rm l}'({\bf u}) \gamma_5 \zeta_{\rm b}'({\bf v}) \,
    \bar \zeta_{\rm b} ({\bf y}) \gamma_5 \zeta_{\rm l} ({\bf z}) \ra, \\
\kone &=& -Z_\zeta^4 \,{a^{12} \over 6L^6} \, \sum_{{\bf u}, {\bf v}, {\bf y}, {\bf z}, k}
\la \bar \zeta_{\rm l}'({\bf u}) \gamma_k \zeta_{\rm b}'({\bf v}) \,
    \bar \zeta_{\rm b} ({\bf y}) \gamma_k \zeta_{\rm l} ({\bf z}) \ra .
\eea
Expanding these correlators at the $\minv$ order of HQET, and using
spin-flavor symmetry, one finds~\footnote{The reader can find the definitions
of the various correlators $f^{\rm stat}_{{\rm A}, 1},f^{\rm kin}_{{\rm A}, 1}, 
f^{\rm spin}_{{\rm A}, 1}, \fdeltaastat$ in~\cite{DellaMorte:2006cb}.} 
\bea
\label{eq:fa1m}
\fa (x_0) &=& 
\zahqet \zzetah\zzeta \rme^{-\mhbare x_0}
\left\{ \fastat(x_0) + \cahqet \fdeltaastat(x_0) + \omegakin \fakin(x_0)
+ \omegaspin \faspin(x_0)
\right\}\,, \nn \\
\kv (x_0)&=& 
-\zvhqet  \zzetah\zzeta \rme^{-\mhbare x_0}
\left\{ \fastat(x_0) + \cvhqet \fdeltaastat(x_0) + \omegakin \fakin(x_0)
-\frac13 \omegaspin \faspin(x_0) \right\}\,, \nn \\
\label{eq:fone1m}
\fone &=& 
\zzetah^2\zzeta^2 \rme^{-\mhbare T}
\left\{ \fonestat + \omegakin \fonekin + \omegaspin \fonespin \right\}\,, \nn \\
\label{e:koneexp}
\kone &=& 
\zzetah^2\zzeta^2 \rme^{-\mhbare T}
\left\{ \fonestat + \omegakin \fonekin
-\frac13 \omegaspin \fonespin
\right\}\,,\nn
\eea
where $\mhbare$ is the (linearly divergent) bare quark mass.
\subsection{Basic observable}
We consider a volume $L^3$ with a time extent $T=L$, and define (in QCD) 
\be
\label{eq:PhiBQCD}
\Phi_{\rm F}^{\rm QCD}(L) = 
\ln \left(  {{-f_{\rm A}(L/2)}\over{\sqrt{f_1}}}
\right)
\nn
\ee
In the large volume limit, 
this observable is related to the decay constant, $F_{\rm B}$, by 
\be
\Phi_{\rm F}^{\rm QCD}(L) 
\stackrel{L \gg 1/\Lambda} \longrightarrow
\ln \left( \frac{1}{2}  F_{\rm B} \sqrt{m_{\rm B}L^3}\right) .
\ee
In a small volume~\footnote{The matching is done in a small volume,
in order to be able to simulate a b-quark with the discretization errors
under control. }
of space extent $L_1\simeq 0.4 \, \fm$, this observable is matched to 
its HQET expression
\be
\Phi_{\rm F}^{\rm QCD}(L_1) = \Phi_{\rm F}^{\rm HQET}(L_1) \; .
\ee
Using the expansions of the correlators $\fa$ and $\fone$ given previously,  
one finds for the rhs
at the static and at the $\minv$ order
\bea
\Phi_{\rm F}^{\rm stat}(L) &=& \ln \zastat 
+ { \ln \left(  {-\fastat(L/2) \over \sqrt{\fonestat}} 
\right)} 
+{\rm O}(\minv),
\\
\label{eq:PhiBhqet}
\Phi_{\rm F}^{\rm stat}(L) + \Phi_{\rm F}^{\minv}(L) &=& \ln \zahqet 
+ 
\ln \left(  {-\fastat(L/2) \over \sqrt{\fonestat}} \right)  
+ \cahqet {\fdeltaastat(L/2) \over \fastat(L/2)}
\\
&&
+ \omegakin \left(  {\fakin(L/2)  \over \fastat(L/2)} 
 - {\fonekin \over \fonestat} \right)
+ \omegaspin \left( {\faspin(L/2) \over \fastat(L/2)}
 - {\fonespin \over \fonestat} \right)
+{\rm O}(\minv^2).\nn
\eea
\subsection{Evolution to larger volumes, in the static approximation}
In order to clarify the discussion, we first explain the strategy in the static
approximation, the generalization to the $\minv$ order will be done in the
next section. 
We start by the matching of HQET to QCD in the volume $L_1$, at the static
order :
$
\Phi_{\rm F}^{\rm QCD}(L_1) = \Phi_{\rm F}^{\rm stat}(L_1).
$
The evolution to a volume $L_{\infty}=L_3=2L_2=4L_1$ is then done, within the effective
theory, in the following way:
\be
\label{eq:Phistat}
\Phi_{\rm F}(L_\infty) =
    [\Phi_{\rm F}^{\rm stat}(L_\infty) - \Phi_{\rm F}^{\rm stat}(L_2)] 
    + [\Phi_{\rm F}^{\rm stat}(L_2) - \Phi_{\rm F}^{\rm stat}(L_1)]
    + \Phi_{\rm F}^{\rm QCD}(L_1) \;.
\ee
We note that $\zastat$ cancels in the differences 
$[\Phi_{\rm F}^{\rm stat}(2L)-\Phi_{\rm F}^{\rm stat}(L)]$.
Using the renormalized SF coupling 
$\bar g^2(L)$~\cite{Capitani:1998mq}, 
we define the static step scaling function (ssf) 
\bea
\sigma_{\rm F}^{\rm stat}(u) 
&=& 
\left[ 
\Phi_{\rm F}^{\rm stat}(2L) - \Phi_{\rm F}^{\rm stat}(L) 
\right]_{\bar g^2(L) = u} 
= \lim_{a/L \to 0} \Sigma_{\rm F}^{\rm stat}(u,a/L) \\ 
\Sigma_{\rm F}^{\rm stat}(u,a/L)
&=&
\label{eq:Sigmazeta}
\left[ 
  \zeta^{\rm stat}(2L) - \zeta^{\rm stat}(L) 
    \right]_{\bar g^2(L) = u} 
\; \mbox{ where }  
\zeta^{\rm stat}(L) =
\ln \left( {-\fastat(L/2) \over \sqrt{\fonestat}} \right)
 \;.
\eea
We can now rewrite the rhs of eq~(\ref{eq:Phistat}) 
as the sum of three continuum terms
\be
\label{eq:Phistat2}
\Phi_{\rm F}(L_\infty) =
\sigma_{\rm F}^{\rm stat}(u_2) + \sigma_{\rm F}^{\rm stat}(u_1) 
+ \Phi_{\rm F}^{\rm QCD}(L_1)\,, \mbox{ where } 
u_{\rm k} = \bar g^2(L_{\rm k}) \;.
\ee
Before discussing the $\minv$ corrections we close this section by a few
remarks :
\begin{itemize}
\item 
The different terms in the eq~(\ref{eq:Sigmazeta}) have to be computed 
at the same value of the lattice spacing (in order to insure that
$\Sigma_F$ has a well defined continuum limit).
This is due to divergences proportional to the logarithm of the lattice spacing
that one has to cancel.
\item In eq.~(\ref{eq:Phistat2}), the entire quark mass dependence comes from 
$\Phi_{\rm F}^{\rm QCD}(L_1)$.
\item At this order, since there is only one matching constant to eliminate 
($Z_{\rm A}^{\rm stat}$), it is sufficient to match one observable ($\Phi_{\rm F}$).
\end{itemize}
\subsection{Including $\minv$ corrections}
At this order, there are three more matching parameters in $\Phi_{\rm F}$ compared
to the static case~\footnote{Also $\zahqet$ is different than $\zastat$, but
as in the static case, it drops out in the differences}.
Therefore, to determine them, we introduce three other
observables defined in a volume of space extent $L$ 
and we give their expressions at the $\minv$ order
\bea
\Phi_1(L) 
\label{eq:Phi1QCD}
&\equiv& \frac{1}{4} (R_1^{\rm P} + 3 R_1^{\rm V}) - R_1^{\rm stat}  
=
\omegakin R_1^{\rm kin} \;,\\
\Phi_2(L)
\label{eq:Phi2QCD}
&\equiv& {{3}\over{4}} \ln \left({{f_1}\over{k_1}}\right) 
=
\omegaspin  {{f_1^{\rm spin}}\over{f_1^{\rm stat}}} 
\qquad \mbox{with } T=L/2
 \;, \\
\Phi_3 (L)
\label{eq:Phi3QCD}
&\equiv& R_{\rm A}(L/2) -R_{\rm A}^{\rm stat}(L/2) 
=
c_{\rm A}^{\rm HQET} R_{\delta {\rm A}}(L/2) 
 + \omegakin R_{\rm A}^{\rm kin}  (L/2)
 + \omega_{\rm spin} R_{\rm A}^{\rm spin}(L/2)  \;,\;
\eea
where the definitions of the ratios $R$ can be found
in~\cite{DellaMorte:2006cb}~\footnote{
We remind the reader that the quantities defined with an subscript 1
are ``boundary to boundary'' observables. Because the noise over signal ratio
grows exponentially with the time, we impose $T=L/2$ for all these
observables.}. 
Together with $\Phi_{\rm F}$, given at this order by~(\ref{eq:PhiBhqet}), we then have a
set of four observables.
In these observables, we have chosen to subtract the static part
(when existing) from the QCD one, as we did in the mentioned reference.
This is perfectly legitimate because they both have a continuum limit,
and this simplifies the equations.\\
Like in the static case, the matching 
is imposed in the volume $L_1$. This allows us to replace in~(\ref{eq:PhiBhqet})
the parameters  $\omegakin$, $\omegaspin$ and $\cahqet$ by a combination of
QCD and HQET quantities. \\
The evolution to the volume $L_2 $ is given by 
\bea
\Phi_{\rm F}(L_2) 
&=& 
\Phi_{\rm F}^{\rm HQET}(L_2) - \Phi_{\rm F}^{\rm HQET}(L_1) + \Phi_{\rm F}^{\rm QCD}(L_1) \\
&=& 
\label{eq:PhiBL2}
\left[\Phi_{\rm F}^{\rm stat}(L_2) - \Phi_{\rm F}^{\rm stat}(L_1)\right]
+
\left[\Phi_{\rm F}^{\minv}(L_2) - \Phi_{\rm F}^{\minv}(L_1)\right] 
+ 
\Phi_{\rm F}^{\rm QCD}(L_1) \;.
\eea
The ssf for the static term has already been given 
in the previous part, and for the $\minv$ part we write
\be
\label{eq:ssf1M}
\Phi_{\rm F}^{\minv}(2L) - \Phi_{\rm F}^{\minv}(L)
=\sum_{\rm i} \sigma_{\rm i}(\bar g^2(L)) \Phi_{\rm i}(L) \; .
\ee
The expressions for the ssf can be found from the 
last equation by using (\ref{eq:PhiBhqet}) together 
with~(\ref{eq:Phi1QCD}), (\ref{eq:Phi2QCD}), (\ref{eq:Phi3QCD}) in the
volume $L_1$. The explicit definitions are given in the appendix.\\
In the step $L_2\rightarrow L_\infty$, we need $\Phi_{\rm i}(L_2,M)$,
and we are then lead to define the ssf for  the $\Phi_{\rm i}$ :
\be
\label{eq:ssf1Mij}
\Phi_{\rm i}(L_2) 
= \sum_j \sigma_{\rm ij}(\bar g^2(L)) \Phi_{\rm j}(L_1) \, .
\ee
We can write down the final equation for $\Phi_{\rm F}$
\be
\label{eq:final}
\Phi_{\rm F}(L_\infty) 
=  \sigma_{\rm F}^{\rm stat}(u_2) + \sigma_{\rm F}^{\rm stat}(u_1)
+ \sum_{\rm ij} \sigma_{\rm i}(u_2) \sigma_{\rm ij}(u_1) \Phi_{\rm j}(L_1)
+ \sum_{\rm i} \sigma_{\rm i}(u_1) \Phi_{\rm i}(L_1)
+\Phi_{\rm F}^{\rm  QCD}(L_1)
\ee
\section{Results in the quenched approximation}
\begin{figure}[!ht]
\begin{center}
\begin{tabular}{cc}
\includegraphics[width=6cm]{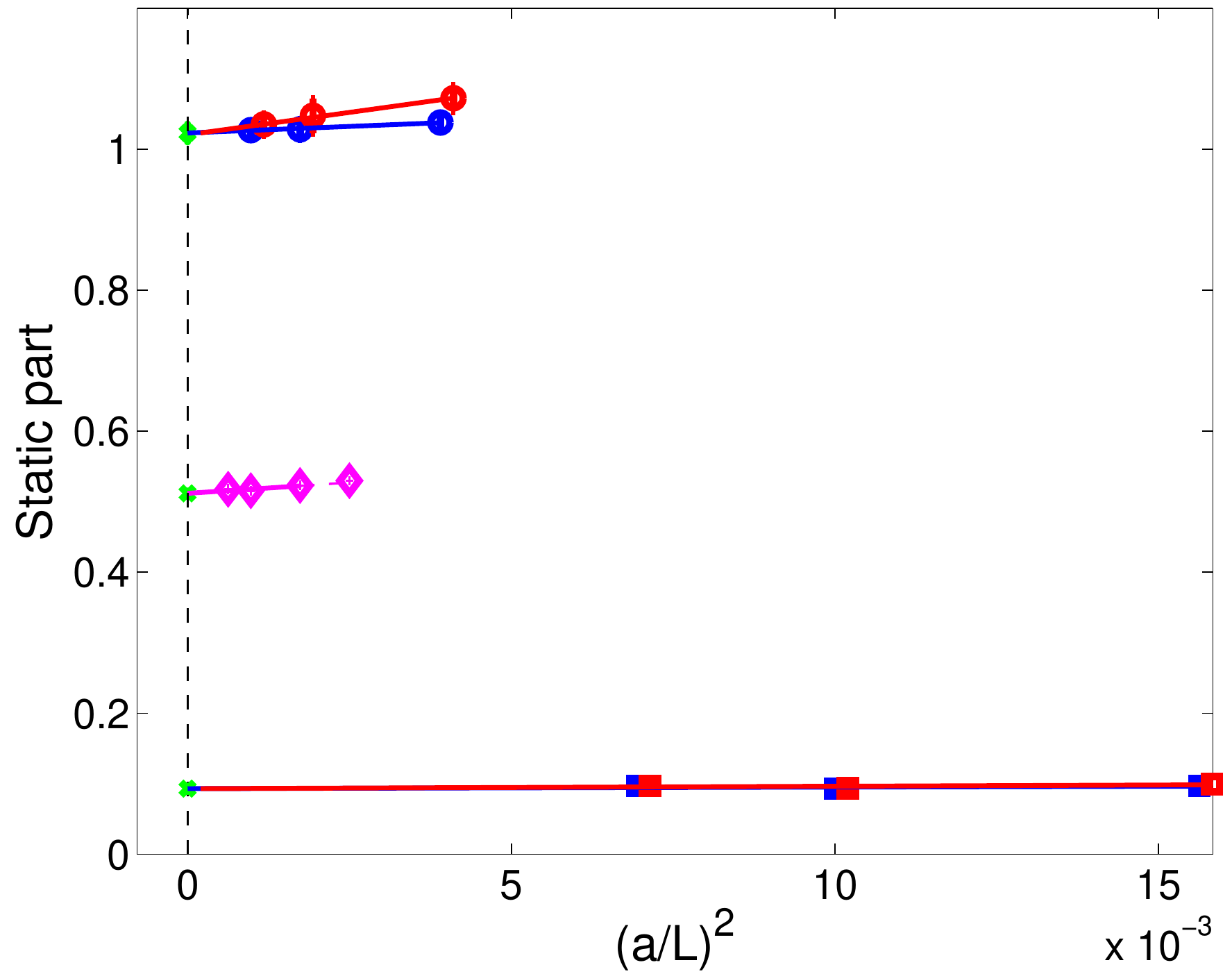} &
\includegraphics[width=6cm]{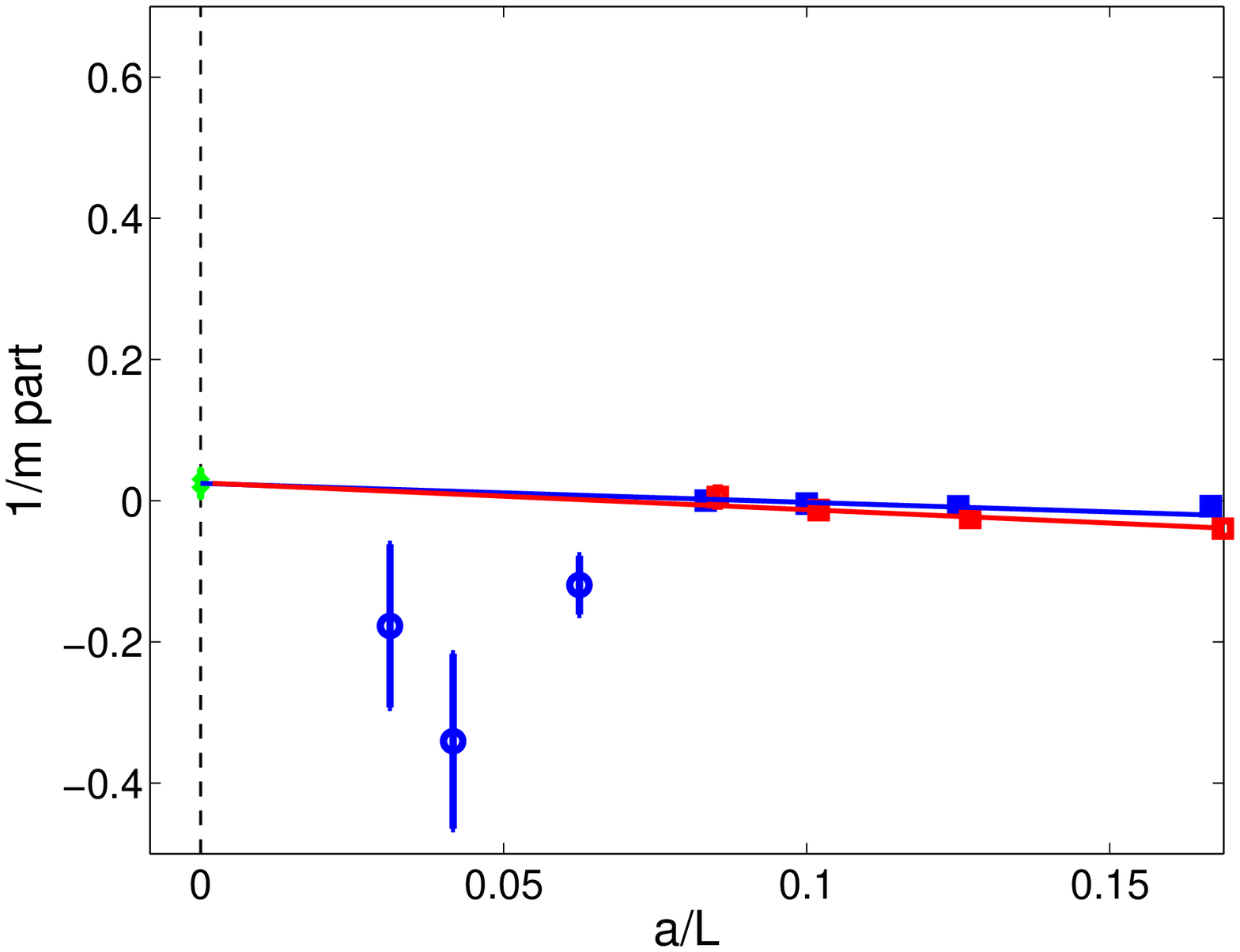} 
\end{tabular}
\end{center}
\caption[]{Continuum extrapolations of the various terms appearing in 
eq.~(\ref{eq:final}). The QCD contribution $\Phi_{\rm F}^{\rm QCD}$ (diamonds) 
and the static part are shown on the left (the circles represent 
the large volume contribution $\sigma_{\rm F}^{\rm stat}(u_2)$, 
and the squares the small
volume part $\sigma_{\rm F}^{\rm stat}(u_1)$ ). The $\minv$ correction is plotted
on the right, the circles represent the large volume terms 
$\sum_{\rm ij} \sigma_{\rm i}(u_2) \sigma_{\rm ij}(u_1) \Phi_{\rm j}(L_1)$,
and the squares the small volume part 
$\sum_{\rm i} \sigma_{\rm i}(u_1) \Phi_{\rm i}(L_1)$.
The color blue stands for HYP2, and red for HYP1~\cite{DellaMorte:2005yc}.}
\label{fig:cont_extr}
\end{figure}

We used basically the same data as in~\cite{DellaMorte:2006cb}, in which the
reader can find the details of the simulation. The simulations are done with
non-perturbatively $\rm O(a)$-improved Wilson fermions. 
The light quark mass is fixed to the strange
one. Concerning HQET, 
we used the HYP actions~\cite{DellaMorte:2005yc}, which help to have
a reasonable statistical precision. 
We show the continuum extrapolations in fig.~\ref{fig:cont_extr}.\\
The extrapolations are done linearly in $(a/L)^2$ for QCD as well as 
for the static part, but in $a/L$ for the $1/m$ term, 
because of the absence of $\rm O(a)$ improvement.
The ordinate scale is the same in order to compare the relative size of the 
different contributions. 
Concerning the precision, one can see that the total error is largely
dominated by the one of the $1/m$ part in the large volume. 
Since for this part 
the results are not yet completely satisfactory, 
we refrain from performing a continuum extrapolation.
We will use the result at the finest lattice spacing only 
($\beta\sim 6.45,\, L/a = 32$).\\
Our preliminary results for $F_{B_s}$ are shown in table~\ref{table_F_Bs_MeV}. 
In the first column we give the results in the static approximation, while in
the other columns, we have included the $\minv$ corrections. 
We observe that in the static approximation, depending on the matching
condition represented here by $\theta$~\footnote{The quark fields are periodic
in space up to a phase $\theta$.} , the result can change by $7\%$. 
This variation disappears when the $\minv$ terms are included. 
Note that differences of $F_{\rm B_{\rm s}}^{\rm stat + \minv}$, 
table~\ref{table_F_Bs_MeV}, have much smaller error than their individual
values,
for example  
\be
\label{eq:diff}
 F_{\rm B_{\rm s}}^{\rm stat + \minv}(\theta_0=0, \theta_1=1, \theta_2=0 )
-F_{\rm B_{\rm s}}^{\rm stat + \minv}(\theta_0=1, \theta_1=0, \theta_2=0.5 )
=4\pm 2 \mbox{ MeV}.
\ee
The other information is that the
$\minv$ term contributes (with a minus sign) up to $\sim 15 \%$ to the final result. 
One can see that adding the $\minv$ terms increases the size of the
statistical errors, as expected from the previous plots.
This is due to the fact that the signal for the $\minv$ 
part in large volume is more
difficult to extract than in the static case, and also because of the absence of
$\rm O(a)$-improvement at this order. 
We also note that our result is compatible with a recent computation done
with a different method, but which also goes beyond the leading order of
HQET~\cite{Guazzini:2006bn}.

\input{table_FB}

\section{Conclusion}
We have shown how to perform a non-perturbative computation of a heavy-light
decay constant at $\minv$ order of HQET, and we have given 
preliminary numerical results in the quenched approximation.
The inclusion of the dynamical quarks is on the 
way~\cite{Jochen:2007,Rainer:2007}.
Applying this method for $N_f>0$ should allow for precise
computations of the heavy-light decay constant, with a good control on the
systematic errors. 
Note in particular that eq.~(\ref{eq:diff}) is a good sign of the absence of
significant $\minv^2$ corrections.
On the numerical side, the cancellations of the 
divergences require sufficient statistical precision, 
and we hope that the all-to-all propagator, like proposed in
\cite{Foley:2005ac,Luscher:2007se}
will be of great help there.\\*
\\
{\bf Acknowledments}\\
We thank NIC for allocating computer time on the APE computers to this project and the
APE group at Zeuthen for its support.

\section{Appendix: The step scaling functions}

In order to have more compact notations, we 
replace $\Phi_{\rm F}$ by 
$\Phi_4$, such that the ssf $\sigma_{\rm i}$ introduced in eq.~(\ref{eq:ssf1M})
are now represented by $\sigma_{\rm 4i}\,$. We can rewrite
eq.~(\ref{eq:ssf1M}) together with eq.(~\ref{eq:ssf1Mij}) as
\be
\Phi_{\rm i}(L_{\rm k+1},M) 
= \sum_{j=1}^4 \sigma_{\rm ij}(u_{\rm k}) \Phi_{\rm j}(L_{\rm k},M) 
+ \delta_{\rm 4i}\sigma_{\zeta}(u_{\rm k})
\, .
\ee
The ssf are then given by a four by four matrix 
$$
\left[\sigma_{\rm ij}\right]
=
\left(
\begin{tabular}{cccc}
$\sigma_{11}$ &    0          &  0            & 0 \\
     0        & $\sigma_{22}$ &  0            & 0 \\
$\sigma_{31}$ & $\sigma_{32}$ & $\sigma_{33}$ & 0 \\
$\sigma_{41}$ & $\sigma_{42}$ & $\sigma_{43}$ & 1 \\
\end{tabular}
\right)
$$
\\
To give their explicit expressions, we define 
$$
\Psi^{\rm kin}(L)
= 
\left( {\fakin (T/2) \over \fastat(T/2)} - {\fonekin \over \fonestat} \right)\;,
\quad
\Psi^{\rm spin}(L)
= 
\left( {\faspin (T/2) \over \fastat(T/2)} - {\fonespin \over \fonestat} \right)
\quad
\mbox{and}
\quad
\rho_{\delta_{\rm A}}(L)= {\fdeltaastat (T/2)\over \fastat(T/2)}  \;.\nn
$$
Then, one finds:
\bea
\Sigma_{11}(u) 
&=&
\left[ R_1^{\rm kin}(2L) / R_1^{\rm kin}(L) 
\right]_{\bar g^2(L)=u} \;,
\qquad 
\Sigma_{22}(u) 
=
\left[ \rho_1^{\rm spin}(2L) / \rho_1^{\rm spin}(L) 
\right]_{\bar g^2(L)=u}  \;,
\nn \\
\Sigma_{31}(u) 
&=&
\left[ { 1 \over R_1^{\rm kin}(L) } 
\left(  R_{\rm A}^{\rm kin}(2L) 
-  {R_{\rm A}^{\rm kin}(L) \, R_{\delta_{\rm A}}(2L) 
\over R_{\delta_{\rm A}}(L)}
\right)
\right]_{\bar g^2(L)=u}   \;,
\nn\\
\Sigma_{32}(u) 
&=&
\left[ { 1 \over \rho_1^{\rm spin}(L) } 
\left(  R_{\rm A}^{\rm spin}(2L) 
-  {R_{\rm A}^{\rm spin}(L) R_{\delta_{\rm A}}(2L) 
\over R_{\delta_{\rm A}}(L)}
\right)
\right]_{\bar g^2(L)=u}  \;,
\qquad
\Sigma_{33}(u)
=
\left[
{R_{\delta_{\rm A}}(2L) \over R_{\delta_{\rm A}}(L)}
\right]_{\bar g^2(L)=u}  \nn  \;,
\\
\Sigma_{41}(u) 
&=& \left[ {1 \over R_1^{\rm kin}(L)} 
\left( \Psi^{\rm kin}(2L) -  \Psi^{\rm kin}(L) 
-{\rho_{\delta_{\rm A}}(2L) - \rho_{\delta_{\rm A}}(L)
\over R_{\delta_{\rm A}}(L)} \, R_{\rm A}^{\rm kin}(L)
\right) \right]_{\bar g^2(L)=u}   \;,
\nn  \\
\Sigma_{42}(u) 
&=& \left[{1 \over \rho_1^{\rm spin}(L)} 
\left( \Psi^{\rm spin}(2L) -  \Psi^{\rm spin}(L) 
-{\rho_{\delta_{\rm A}}(2L) - \rho_{\delta_{\rm A}}(L)
\over R_{\delta_{\rm A}}(L)} \, R_{\rm A}^{\rm spin}(L)
\right) \right]_{\bar g^2(L)=u}   \;,
\nn \\
\Sigma_{43}(u) 
&=& 
\left[ 
{\rho_{\delta_{\rm A}}(2L) - \rho_{\delta_{\rm A}}(L)
\over R_{\delta_{\rm A}}(L)}
\right]_{\bar g^2(L)=u}   \;.
\nn
\eea
\bibliography{refs}  
\bibliographystyle{JHEP}
\end{document}

%% file: table_FB.tex
 \begin{table}[!htb] 
 \hspace{-1.cm} 
 \begin{center} 
 \begin{tabular}{||c||c||c|c|c||}
 \hline
 \hline
 &  $F_{\rm B_s}^{\rm stat}$  &  \multicolumn{3}{c||}{$F_{\rm B_s}^{\rm stat+\minv}$}  \\
 \hline
 \hline
 $\theta_0 $  &   & $\theta_1 = 0$ & $\theta_1 = 0.5$ & $\theta_1 = 1$ \\
              &   & $\theta_2 = 0.5$ & $\theta_2 = 1$ & $\theta_2 = 0$ \\
\hline
 $0$   &  $224 \pm 3$  & $185 \pm 21$  &  $186 \pm 22$  &  $189 \pm 22$ \\
\hline
 $0.5$ &  $220 \pm 3$  & $185 \pm 21$  &  $187 \pm 22$  &  $189 \pm 22$ \\
\hline
 $1$   &  $209 \pm 3$  & $184 \pm 21$  &  $185 \pm 21$  &  $188 \pm 22$ \\
 \hline
 \hline 
 \end{tabular} 
 \end{center} 
 \caption[ ]{Results for $F_{B_s}$ in MeV
with and without the $\minv$ corrections, 
for different  values of the $\theta$ angles.}
 \label{table_F_Bs_MeV} 
 \end{table}